# Fatigue-free ferroelectricity in Hf$_{0.5}$Zr$_{0.5}$O$_2$ ultrathin films via interfacial design

**Authors**

Chao Zhou[1,†], Yanpeng Feng[2,†], Liyang Ma[3,†], Haoliang Huang[4], Yangyang Si[1], Hailin Wang[1], Sizhe Huang[1], Jingxuan Li[1], Chang-Yang Kuo[5], Sujit Das[6], Yunlong Tang[2,7,*], Shi Liu[3,*], Zuhuang Chen[1,*]

**Affiliations**

[1]State Key Laboratory of Advanced Welding and Joining of Materials and Structures, School of Materials Science and Engineering, Harbin Institute of Technology, Shenzhen, 518055, China.

[2]Shenyang National Laboratory for Materials Science, Institute of Metal Research, Chinese Academy of Sciences, Wenhua Road 72, Shenyang, 110016, China.

[3]Key Laboratory for Quantum Materials of Zhejiang Province, Department of Physics, School of Science, Westlake University, Hangzhou, Zhejiang 310024, China.

[4]Quantum Science Center of Guangdong-Hong Kong-Macao Greater Bay Area, Shenzhen 518045, China.

[5]Department of Electrophysics, National Yang Ming Chiao Tung University, Hsinchu, 30010 Taiwan, China.

[6]Material Research Centre, Indian Institute of Science, Bangalore 560012, India.

[7]School of Materials Science and Engineering, University of Science and Technology of China, Shenyang 110016, China

*Corresponding author. Email: zuhuang@hit.edu.cn; liushi@westlake.edu.cn; yltang@imr.ac.cn

†These authors contributed equally to this work

**Abstract**

Due to traits of CMOS compatibility and scalability, $HfO_2$-based ferroelectrics are promising candidates for next-generation memory devices. However, their commercialization has been greatly hindered by reliability issues, with fatigue being a major impediment. We report the fatigue-free behavior in interface-designed $Hf_{0.5}Zr_{0.5}O_2$-based heterostructures. A coherent $CeO_{2-x}$/$Hf_{0.5}Zr_{0.5}O_2$ heterointerface is constructed, wherein $CeO_{2-x}$ acts as an "oxygen sponge", capable of reversibly accepting and releasing oxygen vacancies. This design effectively alleviates defect aggregation at the electrode-ferroelectric interface, enabling improved switching characteristics. Further, a symmetric capacitor architecture is designed to minimize the imprint, thereby suppressing the cycling-induced oriented defect drift. The two-pronged technique mitigates oxygen-voltammetry-generated chemical/energy fluctuations, suppressing the formation of paraelectric phase and polarization degradation. The design ensures a fatigue-free feature exceeding $10^{11}$ switching cycles and an endurance lifetime surpassing $10^{12}$ cycles for $Hf_{0.5}Zr_{0.5}O_2$-based capacitors, along with excellent temperature stability and retention. These findings pave the way for developing ultra-stable hafnia-based ferroelectric devices.

**One-Sentence Summary:** Interface-designed $CeO_{2-x}$/$Hf_{0.5}Zr_{0.5}O_2$ ultrathin heterostructures exhibit fatigue-free behavior exceeding $10^{11}$ switching cycles.

## Introduction

Ferroelectric memories elicit great interests for the merits of reversible polarization states(*1*), nonvolatile characteristic(*1*), fast switching speed(*2*) and low energy consumption(*3*), all of which cater to the demands of next-generation low-power memory and logic devices, such as ferroelectric random access memory(*3*, *4*), ferroelectric field-effect transistors(*5–7*), as well as emerging computing paradigms like neuromorphic and in-memory computing(*8*). The exceptional ferroelectricity discovered in the high-$\kappa$ dielectric $HfO_2$-based ultrathin films(*9*), which are fully compatible with Si-technologies and overcome the thickness limitation(*10*, *11*), provides a solution to tackle the CMOS compatibility and scalability concerns inherent in conventional perovskite oxide ferroelectrics(*12*, *13*). Since its discovery, $HfO_2$-based ferroelectrics have greatly facilitated the integration of ferroelectricity into integrated circuits and boosted the progress of ferroelectric memories(*14*). Despite that, reliability characteristics of $HfO_2$-based ferroelectric electronics, particularly fatigue resistance(*4*, *15*), fall short of commercial metrics(*16*). It is well established that a stable polarization state is fundamental to the operation of ferroelectric devices. Nevertheless, fatigue, characterized by a decrease in switchable polarization during repeated read-write operations, renders polarization states indiscernible and ultimately leads to device malfunction. Till now, most $HfO_2$-based planar capacitor devices can only sustain a stable cycling behavior for ~$10^6$ cycles(*17–20*). This is considerably lower than those of perovskite oxide counterparts (such as $SrBi_2Ta_2O_9$(*21*), $Bi_{3.25}La_{0.75}Ti_3O_{12}$(*22*) and $PbZr_xTi_{1-x}O_3$(*23*, *24*)), which can bear steady performances for >$10^{10}$ switching cycles(*23*, *25*). Additionally, other reliability-related concerns, such as endurance, retention and temperature stability, continue to pose obstacles[4]. Altogether, these reliability issues present substantial impediments in advancing high-performance $HfO_2$-based ferroelectric devices.

Several models have been proposed to explain the fatigue degradation mechanism(*17*, *26*, *27*). Field-driven defect generation and migration, particularly of oxygen vacancies ($V_O$), along with consequent defect trapping and accumulation, are often attributed to these performance instabilities. Especially, several unique features inherent in the unconventional polar phase of $HfO_2$ likely cause its increased susceptibility to fatigue. The ferroelectric phase of $HfO_2$ (orthorhombic $Pca2_1$), compared to the ground-state monoclinic $P2_1/c$ paraelectric phase, is metastable, and would transition into the non-polar phase when subjected to disturbances like oxygen voltammetry(*28*) during electrical cycling. Additionally, the high coercivity characteristic of $HfO_2$-based ferroelectrics(*11*) provides sufficient driving force for the generation and migration of $V_O$(*29*), which accelerates fatigue. The ultra-thin thickness of $HfO_2$-based ferroelectric film not only amplifies these effects, but also makes the interfacial-related degradations (like domain pinning and phase transition) more pronounced. These aspects collectively trigger an early fatigue during operations. Strategies like oxygen scavenging(*30*, *31*), oxygen supply(*32*, *33*), and electrical rejuvenation(*34*), aimed at maintaining the metastable polar phase, reducing defect segregation and releasing pinned domain during cycling, have shown some success in enhancing endurance and relieving fatigue. However, cycling behaviors of optimized $HfO_2$-based planar capacitor devices still lag far behind those of commercial perovskite oxide-based ferroelectric devices[4,](*29*). A deeper understanding of underlying mechanisms of fatigue and further optimization of the reliability of $HfO_2$-based ferroelectrics are urgently needed.

Here, we demonstrate a well-designed interface that can drastically improve reliability characteristics of $HfO_2$-based ultrathin ferroelectric devices. An ultrathin fluorite-structured $CeO_{2-x}$ is engineered as a coherent capping layer on ferroelectric $Hf_{0.5}Zr_{0.5}O_2$ (HZO) film, effectively reducing the switching barrier during polarization reversals. Moreover, the multivalent-oxide $CeO_{2-x}$ epitaxial layer not only reduces the diffusion barrier of $V_O$ near the $CeO_{2-x}$/HZO interface region but also serves as an "oxygen-sponge", which can reversibly accept and release $V_O$. This approach mitigates the ferroelectric-to-paraelectric phase transformation and maintains the polar feature of HZO throughout the cycling process, yielding an ultra-thin $CeO_{2-x}$/HZO heterojunction ferroelectric device with notably enhanced polarization ($P_r$ = 21 μC cm$^{-2}$) and reduced coercive field ( < 3 MV cm$^{-1}$), alleviated imprint and stable retention. Furthermore, incorporating an additional LSMO layer on the CeO$_{2-x}$ film as a secondary buffer constructs a more symmetric capacitor (i.e., Pt/LSMO/CeO$_{2-x}$/HZO/LSMO). This design suppresses the imprint field-driven oriented defect movement, enables comprehensive improvement of ferroelectric properties and achieves an ultra-stable cycling behavior of hafnia-based devices, which manifests as a fatigue-free cycling behavior approaching 10$^{11}$ electrical cycles and a stable endurance lifetime surpassing 10$^{12}$ cycles.

## Results

### DFT computational results

It is well known that ferroelectric-electrode heterointerfaces, where defect accumulation(*26*), charge injection(*35*) and domain pinning(*36*, *37*) occur, are crucial to the cycling reliability of ferroelectric materials. These interfacial phenomena would play an important role in the endurance and fatigue behaviors of HfO$_2$-based ferroelectric devices. Therefore, optimizing interfaces is a natural and feasible strategy for enhancing reliability.

We choose fluorite-structured CeO$_{2-x}$ as the optimal interfacial buffer for HfO$_2$ because of its structural similarity to HfO$_2$ and its oxygen-active properties(*38–40*). Our first-principles calculations reveal that the migration barrier for V$_O$ in bulk HfO$_2$ is approximately 2.3 eV (Fig. 1A). Notably, the incorporation of a CeO$_{2-x}$/HfO$_2$ interface activates V$_O$ in HfO$_2$ near the interface region, substantially reducing the V$_O$ diffusion barrier to a range of 0.22-0.96 eV. Therefore, V$_O$ becomes mobile in the interface zone, alleviating the defect pinning and enabling an effortless movement of V$_O$ across the CeO$_{2-x}$/HfO$_2$ heterointerface. The CeO$_{2-x}$ buffer, by promoting interfacial vacancy mobility, effectively serves as an "oxygen sponge", rendering the reversible storage and release of V$_O$ during electrical read-write operations. As schematized in Fig. 1B, this oxygen-active CeO$_{2-x}$/HfO$_2$ heterointerface facilitates the timely transportation of V$_O$ under electric fields. In this configuration, the CeO$_{2-x}$ layer, rather than the HfO$_2$ film, acts as the source and drain of V$_O$, preventing the accumulated change of oxygen concentration within HfO$_2$. This design could maintain the metastable polar HfO$_2$ phase and mitigate the early degradation of switching performance of HfO$_2$-based ferroelectric devices. Conversely, other interfaces, such as ZrO$_2$/HfO$_2$ (Supplementary fig. S1) or bare HfO$_2$ (Fig. 1B), are less competent in activating V$_O$. This inefficiency leads to the aggregation of defects near the interface, likely causing domain pinning and phase transition during cycling operations. Furthermore, our DFT calculations suggest that the higher symmetry of the cubic CeO$_{2-x}$ layer could stabilize the tetragonal phase of HfO$_2$, thereby lowering the thermodynamic energy difference between orthorhombic ferroelectric phase and tetragonal paraelectric phase. Consequently, the switching barrier for the CeO$_{2-x}$/HfO$_2$ heterostructure is decreased significantly (Fig. 1C), contributing to a smaller coercive field and enhanced endurance performance.

### Characterizations of capped Hf$_{0.5}$Zr$_{0.5}$O$_2$

In accordance with the computational guidance, a ~6-nm-thick HZO film capped with ~0.8 nm CeO$_{2-x}$ layer was fabricated (detailed in the Methods part). Additionally, HZO films with a capping layer of either ZrO$_2$ or Al$_2$O$_3$ oxides were also prepared for comparisons (Supplementary fig. S2). XRD $\theta$−$2\theta$ scans reveal that all films, grown on (001)-oriented LSMO/SrTiO$_3$, exhibit the (111) orientation of the orthorhombic ferroelectric phase with essentially identical diffraction peak positions (Supplementary fig. S2B). Additionally, clear Laue oscillations near the (111)$_O$ diffraction peak of HZO and smooth film surfaces (Supplementary figs. S2C-F) indicate high crystalline quality and clean interface of the heterostructures. High-angle annular dark-field (HAADF) scanning transmission electron microscopy (STEM) and Electron Energy Loss Spectroscopy (EELS) mappings were further performed to characterize the structure and interface abruptness. An atomic-resolved image of the $CeO_{2-x}$/HZO heterostructure is presented (Fig. 1D), where a thicker $CeO_{2-x}$ capping layer (~1.5 nm) was grown on the HZO to ensure clear observation. HAADF-STEM image demonstrates that the $CeO_{2-x}$/HZO heterointerface is coherent. Further analysis using EELS elemental mapping reveals a distinct interface and minimal interdiffusion within the $CeO_{2-x}$/HZO heterointerface.

Though structurally similar, electrical performances of these heterostructures vary significantly. Compared to bare HZO films, capacitors based on HZO films with $ZrO_2$, $Al_2O_3$ and $CeO_{2-x}$ capping layers exhibit notably lower leakage currents (Supplementary fig. S3A). The differences in polarization switching characteristics are even more pronounced. The positive-up-negative-down (PUND) measurement reveals that all capped HZO devices show enhanced polarizations compared to the bare one (Supplementary figs. S3B-S3C and Fig. 1E). Among them, capacitors based on $CeO_{2-x}$/HZO heterostructures demonstrate the most impressive performance, characterized by the highest remnant polarization ($P_r$ =21 $\mu C$ $cm^{-2}$), the sharpest polarization switching current and the lowest coercive field ($E_C$ =2.9 MV $cm^{-1}$). Conversely, effects of lowering $E_C$ and enhancing $P_r$ in HZO-based devices capped with either $Al_2O_3$ or $ZrO_2$ are less pronounced. These results are consistent with our aforementioned computational predictions. Furthermore, the $CeO_{2-x}$/HZO heterostructure exhibits the highest dielectric constant and most distinct butterfly-shaped $\varepsilon$-$E$ curve (Supplementary fig. S3E).

Benefiting from reduced leakage response, the breakdown strength and endurance capabilities of these capped capacitors have been substantially enhanced (Supplementary fig. S4), in stark contrast to the early breakdown occurring after ~$10^5$ switching cycles in the bare HZO capacitors. However, the fatigue behaviors of $ZrO_2$/HZO and $Al_2O_3$/HZO devices still exhibit an average performance in comparison to reported results(*17*, *18*, *20*). A noticeable degradation occurs in the early stage of cycling for both $ZrO_2$/HZO and $Al_2O_3$/HZO heterostructures, with a drastic decrease in polarization observed after $10^8$ cycles (Supplementary fig. S4E). Switching dynamics characterizations (Supplementary fig. S5) are well consistent with the fatigue measurements of these HZO-based capacitors, manifested as the loss of polarization and the slowdown of switching speed after cycling. This may be attributed to the defect pinning and phase transitions occurring during the electrical cycling. In contrast, the "oxygen sponge" $CeO_{2-x}$ buffer reduces $V_O$ diffusion barrier and decreases coercive field in the $CeO_{2-x}$/HZO device, which jointly ensure the exceptional reliability: the $CeO_{2-x}$/HZO-based capacitor demonstrates robust cycling stability, maintaining stable performance for over $2\times10^9$ cycles and an endurance lifespan exceeding $2\times10^{10}$ cycles (Fig. 1F-1H) as well as a stable dynamics profile (Supplementary fig. S5E). This positions it among the most reliable categories of documented metal-ferroelectric-metal (MFM) planar devices(*17*, *18*, *20*).

**Oriented drift of defects**

To elucidate the impact of the $CeO_{2-x}$ capping layer, semi in-situ micro-area X-ray photoelectron spectroscopy was employed to probe the valence state of $CeO_{2-x}$/HZO before and after electrical poling (Fig. 2A and Supplementary fig. S6). For the as-grown state, the Ce cation within the $CeO_{2-x}$ capping layer exhibits a multivalent oxidation state comprising both $Ce^{3+}$ and $Ce^{4+}$, with the $Ce^{3+}$ peak area constituting a modest ~17% of the total area. Remarkably, the trivalent ($Ce^{3+}$) and tetravalent ($Ce^{4+}$) oxidation states can interconvert under different electrical stimuli. Specifically, the area ratio of $Ce^{3+}$ decreases to ~7% after positive poling, which indicates that a certain amount of $V_O$ migrates from the $CeO_{2-x}$ buffer toward the bottom LSMO electrode. Conversely, a significant amount of $Ce^{4+}$ transitions into $Ce^{3+}$ within the $CeO_{2-x}$ layer under a negative bias. This phenomenon is corroborated by the Hf 4$f$ XPS spectrum (Supplementary fig. S7). Thanks to the protection of a $CeO_{2-x}$ buffer layer, the Hf 4$f$ spectrum exhibits less sensitivity to external bias compared to that of the bare HZO sample. In the case of bare HZO counterpart, the sub-oxide peak emerging after negative pulse treatment (Supplementary fig. S8) exhibits a markedly enhanced intensity compared to that observed in the $CeO_{2-x}$/HZO heterostructure. Collectively, these findings elucidate the "oxygen sponge" characteristic of the $CeO_{2-x}$ layer during electrical operations, highlighting its ability to dynamically regulate $V_O$ concentration within HZO, and thereby enhancing device performance and reliability.

Both computational and experimental results corroborate the reservoir feature of the $CeO_{2-x}$ capping layer, which facilitates the reversible diffusion of $V_O$ in accordance with the orientation of the applied electric field. Ideally, the reversible migration of the $V_O$ would maintain a relatively stable oxygen content within the $CeO_{2-x}$/HZO device, thereby ensuring that the state of the ferroelectric capacitor would not vary significantly during cycling. This is a pivotal factor in sustaining the cycling reliability. Nevertheless, the peak area associated with $Ce^{3+}$ decreases from ~17% to ~13% when the device undergoes $10^8$ switching cycles (Fig. 2A). This signifies a net migration of $V_O$ towards the LSMO bottom electrode and an accumulation of oxygen ions in the $CeO_{2-x}$ capping layer, despite the capacitor being subjected to the symmetric bipolar electric field cycling. This phenomenon is further corroborated by the Ce $M_{4,5}$-edge EELS spectra. Initially, $Ce^{3+}$ ions reside at the $CeO_{2-x}$/HZO heterointerface and extend into the bulk region of the $CeO_{2-x}$ layer (Fig. 2C). After $10^8$ cycles, the $Ce^{3+}$ region becomes narrowed and the signature of $Ce^{4+}$ intensifies (Figs. 2D), supporting the XPS results. The progressively oxygen-rich $CeO_{2-x}$ layer would eventually lose its "oxygen sponge" function, ultimately leading to an inevitable fatigue of the $CeO_{2-x}$/HZO device after $2\times10^9$ cycles.

Synchrotron-based micro-beam XRD studies were further carried out to probe the possible phase transition during the fatigue process. The $CeO_{2-x}$ buffer, functioning as a source and drain for $V_O$ and oxygen ions, prevents the accumulation of $V_O$ or oxygen atoms at the interface. Thus, the HZO layer in the heterostructure retains a stable polar structure even after $10^8$ cycles (Figs 2E-2F). In contrast, for Pt/HZO/LSMO capacitors without the $CeO_{2-x}$ capping layer, due to the absence of a timely transportation of defects, the Pt/HZO interface acts as a terminal for defects and mobile ions, turning the HZO film itself becoming the main "battlefield". This leads to a gradual transformation of the metastable polar structure into the energetically-favored phase due to the field cycling-induced chemical/energy fluctuations. Oxygen accumulation would occur at Pt (top electrode)/HZO interface and induce the formation of the energetically-stable paraelectric monoclinic phase when the bare HZO device cycled $10^6$ times (to breakdown) (Figs. 2G-2H). This observation is consistent with previous study(*28*, *41*). Therefore, the absence of oxygen-active $CeO_{2-x}$ buffer layer leads to structural phase transitions and likely defect pinning at the interface, directly contributing to the fatigue phenomenon.

**Behaviors of more symmetric capacitors**

The oriented net movement of $V_O$ or oxygen ions is usually attributed to the asymmetric electrical potential within the ferroelectric capacitor, leading to the asymmetric polarization and defect accumulation at the interface(*42*). Tracing back to the PUND hysteresis loop of the Pt/$CeO_{2-x}$/HZO/LSMO capacitor (Supplementary fig. S3), we can observe an apparent imprint (i.e., horizontal shift of the *P-E* loop) towards the negative bias, indicating the presence of a positive built-in field ($E_{bi}$), directed from the top electrode towards the bottom. This fixed positive $E_{bi}$ superimposes on the applied external symmetric bipolar field ($\pm E_{ap}$), resulting in an asymmetric electrical potential, manifested as $(+E_{ap} + E_{bi})d$ and $(-E_{ap} + E_{bi})d$ (where $d$ denotes the thickness of the ferroelectric), across the two-terminated capacitor per cycle (schematic illustrated in Fig. 3A). In this scenario, $V_O$ or oxygen ions move back and forth but with different velocities and distances under the electrical driving force with different polarities, triggering the directional migration of $V_O$ or oxygen ions. Herein, Fig. 3B summaries the relationship between the imprint and fatigue behaviors in representative studies (Supplementary Note 1 for references). Interestingly, an empirical trend emerges: the more severe the imprint, the faster the onset of fatigue. This correlation aligns with our prior elucidations that the noticeable imprint provides sufficient driving force for the directional movement of defects.

Fig. 3B further demonstrates that the architecture of a ferroelectric capacitor influences the imprint effect. A symmetric capacitor structure can effectively decrease the imprint effect.

Therefore, it is logical to consider the symmetric "Pt/LSMO/HZO/LSMO" capacitor architecture (Supplementary fig. S9), where the HZO layer is sandwiched between two LSMO films, for improved reliability. Owing to the symmetric capacitor structure, the LSMO-capped HZO specimen does exhibit reduced $E_C$ and smaller imprint (Supplementary fig. S10). These factors contribute to the enhanced cycling performance of the LSMO/HZO/LSMO device, outperforming $ZrO_2$ and $Al_2O_3$-capped counterparts. Despite that, the fatigue behavior of Pt/LSMO/HZO/LSMO capacitor is still inferior to that of the Pt/CeO$_{2-x}$/HZO/LSMO counterpart (Supplementary fig. S10D). We propose that this is probably due to the lower effectiveness of $V_O$ transport in LSMO compared to CeO$_{2-x}$(*43, 44*).

To fully leverage the benefits of the symmetric structure while simultaneously maintaining a reduced $V_O$ interfacial diffusion barrier, we further grow the LSMO layer on top of the CeO$_{2-x}$ (Supplementary fig. S11). The LSMO layer not only serves as a secondary oxygen reservoir but also plays a vital role in minimizing imprint effect. As expected, the ultrathin Pt/LSMO/CeO$_{2-x}$/HZO/LSMO capacitor exhibits a low $E_C$ of 2.34 MV cm$^{-1}$ and suppressed imprint effect (with $E_{bi}$ of only 0.38 MV cm$^{-1}$) (Fig. 3C and Supplementary fig. S12). Accordingly, the oriented drift of defects in the Pt/LSMO/CeO$_{2-x}$/HZO/LSMO capacitor is suppressed compared to that of the Pt/CeO$_{2-x}$/HZO/LSMO capacitor (Supplementary fig. S13). Additionally, the insertion of the LSMO layer on the CeO$_{2-x}$ buffer does not exacerbate the leakage current (Fig. 3D).

The synergistic effects of the reduced $V_O$ migration barrier, the LSMO-CeO$_{2-x}$ bilayer "oxygen sponge" and the mild imprint collectively enable remarkable cycling performance of the Pt/LSMO/CeO$_{2-x}$/HZO/LSMO ferroelectric capacitor, demonstrating virtually fatigue-free behavior for up to $10^{11}$ cycles (the capacitor shows little change in the hysteresis loop and there is only a <5% loss in polarization after the $10^{11}$ switching cycle) and maintaining a stable ferroelectricity even after $10^{12}$ cycles (Figs. 4A-4B and Supplementary figs. S14A-S14B). Meanwhile, pulse switching dynamics measurement (Supplementary fig. S14C) demonstrates a tiny variation of switchable polarization and switching time after field cycling. To our knowledge, this interface-designed device exhibits the most robust stability without requiring any rejuvenation process(*34, 45*) during fatigue measurements, making it the most stable HfO$_2$-based planar capacitor reported to date (Fig. 4C and Supplementary Note 1 for references). Beyond that, the Pt/LSMO/CeO$_{2-x}$/HZO/LSMO capacitors display stable retention exceeding 10 years even when baked at 358 K (Fig. 4D), whereas the bare one loses its ferroelectricity gradually. Distinct from the intense changes in polarization observed in other HZO-based ferroelectric devices(*46, 47*) in varying temperatures, our device exhibits reliable temperature stability (Fig. 4E and Supplementary fig. S15).

**Discussion**

The reliability issue, especially the fatigue failure, is one of the key limitations need to be settled for hafnia-based ferroelectric devices. Here, the oxygen-active and "oxygen sponge" characteristics brought by the CeO$_{2-x}$/HZO heterointerface effectively suppress the defect accumulation and consequent polar-paraelectric phase transition during the field cycling. These enable ferroelectric devices based on CeO$_{2-x}$/HZO heterostructures with enhanced polarization, impressive fatigue-resistance and endurance behavior. Our systematic studies further revealed the detrimental impact of directional defects migration caused by the built-in field (imprint) on fatigue characteristics. By further building a more symmetric-structure capacitor, a record-breaking fatigue-free HfO$_2$-based planar capacitor which can be steadily operated over ~$10^{12}$ cycles is obtained. In addition, the designed device demonstrates a comprehensive improvement of ferroelectric properties which can be defined as the "hexagonal warrior" in view of its attractive traits of fatigue-resistant, retention, coercive field, leakage current, imprint and switchable polarization (Fig. 4F) and the comprehensive reliability performances of this simple designed planar capacitor are even exceeding those of Micron's very recent device with advanced 3D integration and complicated packaging

solutions(*14*). To concluded, through the design of symmetric-structure capacitor with a "oxygen sponge" layer, we have harnessed the migration of $V_O$/oxygen ions during the cycling operation, achieving ideal performances of $HfO_2$-based ferroelectric devices. The method embodied in our work provides a feasible and promising routine to design applicable and high-performance $HfO_2$-based ferroelectric devices.

**Materials and Methods**

**Sample deposition.** Both the LSMO (22 nm)/$CeO_2$ (0.8 nm)/ HZO (6 nm)/LSMO (11 nm) and $CeO_2$ (0.8 nm)/ HZO (6 nm)/LSMO (11 nm) systems were epitaxially grown on (001) $SrTiO_3$ single crystal substrates by pulsed laser deposition (Arrayed Materials RP-B) using a KrF excimer laser ($\lambda = 248$ nm). The HZO layers were deposited at 600 °C under an oxygen partial pressure of 15 Pa at a laser repetition rate of 2 Hz and a laser fluence of 1.35 J cm$^{-2}$. The $CeO_{2-x}$ were fabricated at an oxygen pressure of 10 Pa at 600 °C with a laser fluence of 0.85 J cm$^{-2}$ and a laser repetition rate of 3 Hz. The bottom electrode LSMO layers were grown with a laser (3 Hz, 0.85 J cm$^{-2}$) under an oxygen partial pressure of 20 Pa at the substrate temperature of 700 °C. For the top LSMO capping layers, they were deposited at 600 °C with a laser (3 Hz, 0.85 J cm$^{-2}$) under an oxygen partial pressure of 20 Pa.

**Device fabrication.** All films were processed with a standard lithography to pattern the top electrode mask on films. After photolithography, top electrode Pt layer was deposited on mask areas via magnetron sputtering (Arrayed Materials RS-M). The MFM structure capacitors were fabricated when the photoresist was lifted off. As for LSMO capped devices, the LSMO exposed area that was not covered by Pt was etched by the KI + HCl solution (0.37% HCl + 5 mol L$^{-1}$ KI aqueous solution).

**Micro-area XPS characterizations**. The micro-area X-ray photoelectron spectroscopy was performed using a PHI VersaProbe 4 instrument (Physical Electronics, Inc.). The diameter of the X-ray spot employed in the testing process was 50 $\mu$m, with the apparatus configured to 15 kV and 12.5 W. In the XPS analysis, energy calibration was carried out by using C 1$s$ peak at 284.8 eV.

**X-ray diffraction and topography characterizations.** X-ray $\theta$-$2\theta$ scans were obtained by high-resolution X-ray diffractometer (Rigaku Smartlab 9 KW). For study the phase structure of HZO heterostructures before and after electric treatments, micro-area XRD was carried out at the beamline BL15U1 station in Shanghai Synchrotron Radiation Facility (SSRF). The wavelength and diameter of the X-ray is 1 Å and 5 $\mu$m. The obtained data was explored and processed with the Dioptas programs suit(*48*). The film morphology was determined by an Asylum Research MFP-3D-Infinity atomic force microscopy.

**Aberration-corrected STEM characterization.** The TEM samples for STEM observations were prepared by the focused-ion-beam (FIB) lift-out method. Before cutting, the C and W coating were deposited on the devices to protect the surface of devices. At first, the voltage of 30 kV and current of 0.44 nA for ion beam were used to polish the TEM samples. Then the current was gradually reduced to 41 pA. Finally, the voltage of 5 kV, 2 kV and 1 kV for ion beam were used to clean the surface amorphous. Cross-sectional HAADF-STEM images and EELS mappings were acquired by a double-aberration-corrected scanning transmission electron microscope (Spectra 300, ThermoFisher Scientific) equipped with a monochromator, a Gatan 1069 EELS system and K3 camera operating at 300 kV.

**Electrical characterizations.** Electrical measurements were implemented via a MFM capacitance structure with the bottom electrode LSMO grounded and the top Pt connected to the driving end. The electric performances, including dynamic *P-E* loops, PUND measurements, retention measurements and fatigue behaviors, were measured using a TF 3000 analyzer (aixACCT) and leakage currents was measured using Keithley 4200A-SCS parameter analyzer (Tektronix). All temperature-dependent measurements were carried out in a cryogenic probe station (TTPX, Lake Shore Cryotronics).

**DFT calculations.** All first-principles density functional theory calculations were performed using Vienna Ab initio Simulation Package (VASP) with generalized gradient approximation of the Perdew-Burke-ErnZerhof (PBE) type(*49–51*). To describe the localized 4*f* states of Ce properly, the on-site Coulomb interactions correction was used(*52*). As suggested by previous works, the U value for Ce 4*f* was set to 5.0 eV(*53, 54*). The optimized lattice constants of $CeO_2$ are $a=b=c=5.50$ Å for the cubic phase, in good agreement with the experimental value (5.411 Å). In order to better fit the experimental conditions, we construct a (111)-orientation interface model which contains 12 Ce, 36 Hf and 96 O atoms, corresponding to 3 $CeO_2$ layers and 9 $HfO_2$ layers. The plane-wave cutoff is set to 550 eV. We use a 2×2×1 Monkhorst-Pack *k*-point grid and an energy convergence threshold of $10^{-5}$ eV for structural optimizations. The bottom 6 $HfO_2$ layers (half of $HfO_2$ layers) were fixed to their positions in the defect-free state during structural optimizations and NEB calculations, while the other layers were allowed to relax. We introduce oxygen vacancies at different positions in regions that allow relaxation to determine the migration energy barrier of oxygen vacancies. The minimum energy paths of oxygen vacancy diffusions are determined using the nudged elastic band (NEB) technique with VASP transition state tools (VTST)(*55*). The spring constant was set to 5 eV/Å$^2$, and a force convergence criterion of 0.03 eV/Å was used. The minimum energy paths of polarization switching in $CeO_2$-$HfO_2$ system and pure polar $HfO_2$ are determined using the nudged elastic band (NEB) technique implemented in the USPEX code(*56–58*) with lattice constants clamped during polarization switching. The root-mean-square forces on images smaller than 0.03eV/Å is the halting criteria condition for NEB calculations. The variable elastic constant scheme is used, and the spring constant between the neighboring images is set in the range of 3.0 to 6.0 eV/Å$^2$.

**Acknowledgments**
This work was supported by National Key R&D Program of China (Grant No. 2021YFA1202100), National Natural Science Foundation of China (Grant Nos. 92477129, 52372105 and 12361141821) and Shenzhen Science and Technology Program (Grant No. KQTD20200820113045083). Z.H.C. has been supported by State Key Laboratory of Precision Welding & Joining of Materials and Structures (Grant No. 24-Z-13) and "the Fundamental Research Funds for the Central Universities" (Grant No. HIT.OCEF.2022038). The authors thank the staff from Shanghai Synchrotron Radiation Facility (SSRF) at BL15U1.

**Author contributions:** Z.H.C. and C.Z. conceived and designed the research. C.Z. fabricated the films and carried out electric measurements with the assistance of Y.Y.S., S.Z.H., and J.X.L. C.Z. performed the XRD measurements with the assistance of H.L.H and Y.Y.S. Y.P.F. and Y.L.T. performed the STEM characterizations. L.Y.M. and S.L. carried out theoretical calculations. C.Z., L.Y.M., S.L., S.D., and Z.H.C. wrote the manuscript. Z.H.C. supervised this study. All authors discussed the results and commented the manuscript.

**Competing interests:** Authors declare that they have no competing interests.

**Data and materials availability:** All data are available in the main text or the supplementary materials.

# Figures and Tables

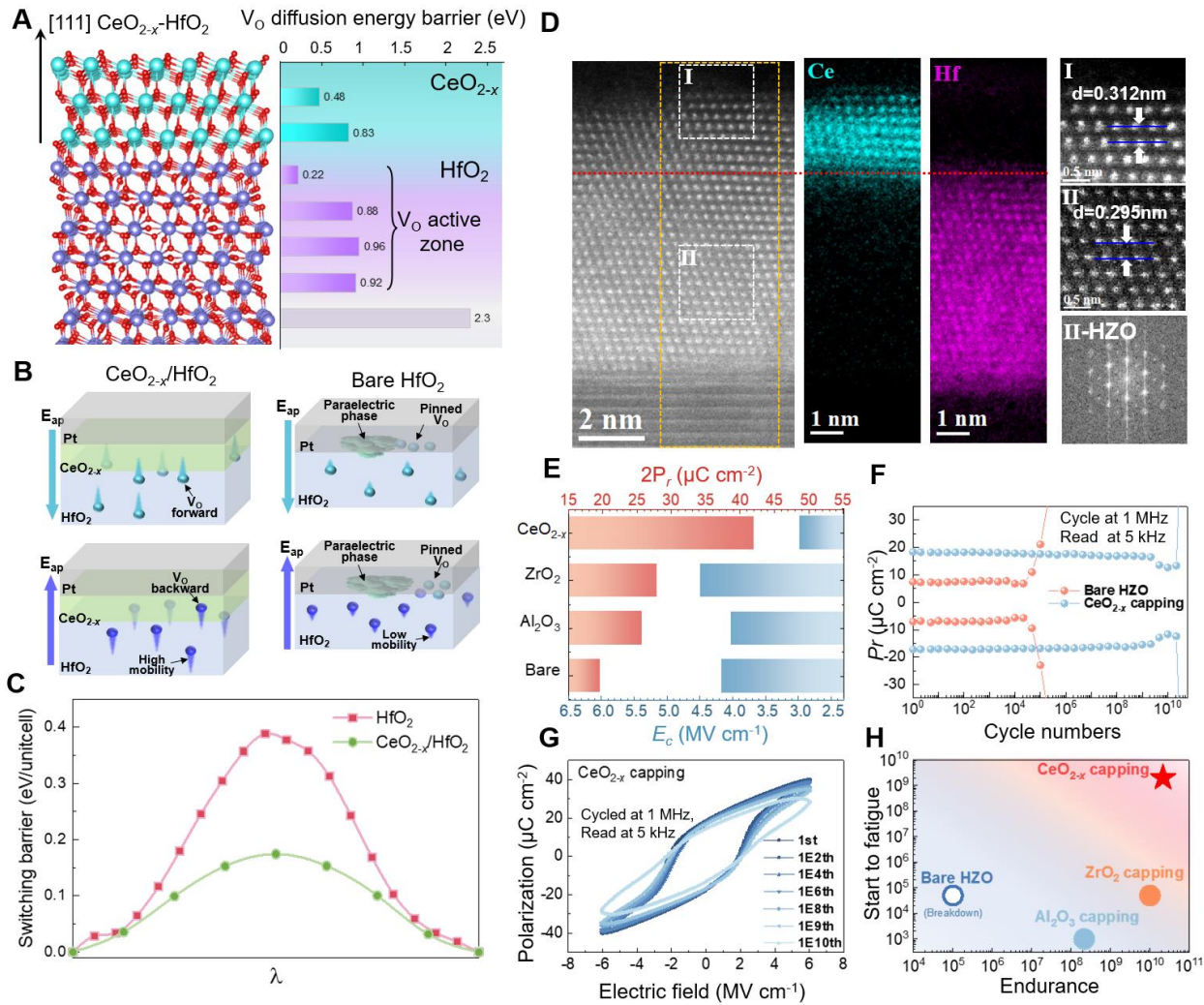

**Fig. 1. Atomic insight and enhanced performances of interface-designed HfO₂-based devices.** (**A**) Diffusion energy barriers of oxygen vacancy ($V_O$) in the $CeO_{2-x}$/$HfO_2$ heterostructure. (**B**) Schematics of $V_O$ migrations in the capacitors based on $CeO_{2-x}$/$HfO_2$ and bare $HfO_2$ under fields. The arrows colored in cyan and blue denote the applied electric fields in different directions. The spheres highlighted in cyan and blue represent the oxygen vacancies moving forward and backward. The length of the tails, irrespective of color, signifies the mobility of defects in the presence of external electric fields. (**C**) Polarization switching barriers for the $CeO_{2-x}$/$HfO_2$ and bare $HfO_2$ capacitors. (**D**) HADDF-STEM image, EELS mappings and Fourier transformation for the $CeO_{2-x}$/$HfO_2$/LSMO heterostructures. (**E**) Coercive fields and remnant polarization values for capped HZO devices. (**F**) Cycling performances for the bare Pt/HZO/LSMO device and Pt/$CeO_{2-x}$/HZO/LSMO capacitor. (**G**) Polarization-electric field loops for the Pt/$CeO_{2-x}$/HZO/LSMO capacitor at different cycling stage. (**H**) A comparison of the fatigue and endurance behaviors for the HZO-based capacitors with different capping layers.

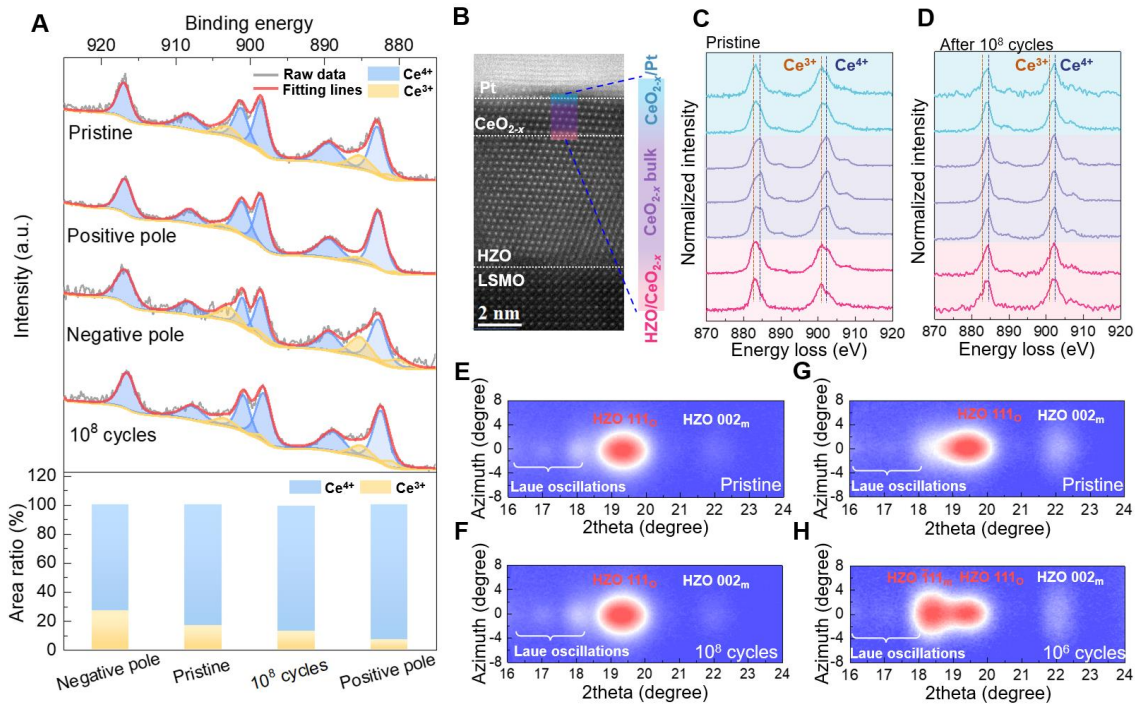

**Fig. 2. Evolution of Ce valence states and structure under different electric treatments.** (**A**) Micro-area X-ray photoelectron spectroscopy (XPS) of the Ce element in the $CeO_{2-x}$/HZO heterostructure after different electric treatments and a summary of $Ce^{3+}$:$Ce^{4+}$ ratio for the corresponding treatments. (**B**)-(**D**) HAADF-STEM image with the marked region for electron energy loss spectroscopy (EELS) measurements, along with the relevant EELS signals before and after endurance cycles. Micro-beam diffraction results for (**E**)-(**F**) the $CeO_{2-x}$/HZO film and (**G**)-(**H**) the bare HZO film before and after endurance cycles. The micro-beam diffractions were performed at the BL15U1 station in Shanghai Synchrotron Radiation Facility with a wavelength of 1 Å.

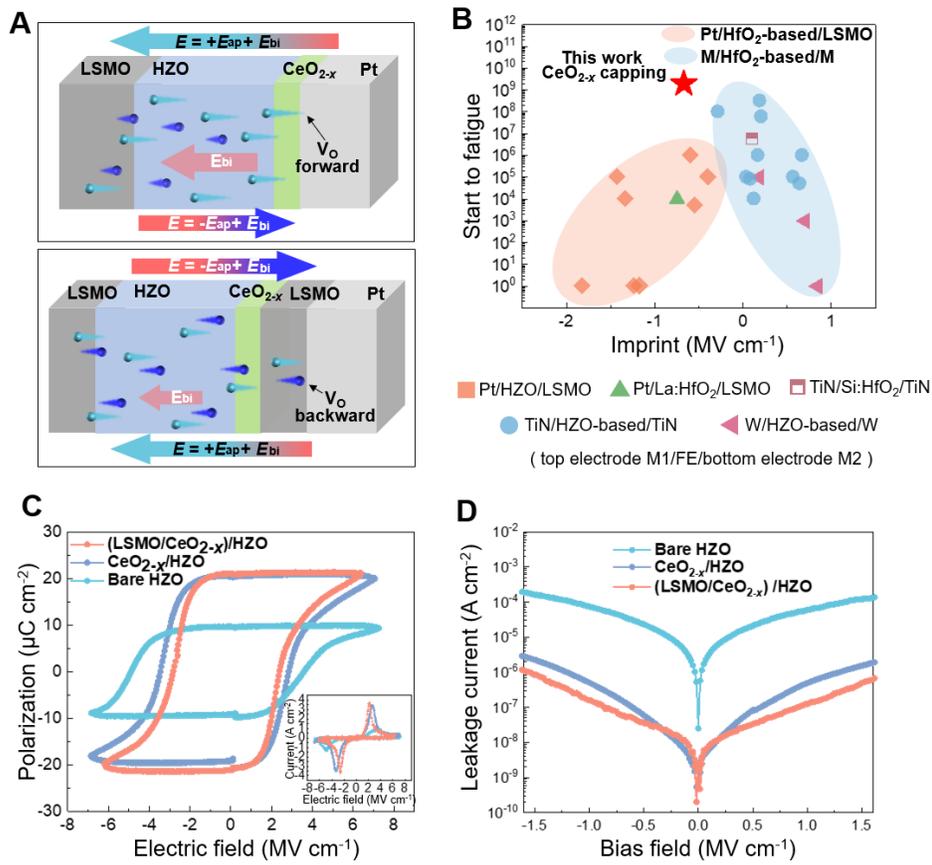

**Fig. 3. Effects of the HZO capacitor architecture on the electric performances.** (**A**) Schematics depicting the impact of capacitor structure on the built-in bias ($E_{bi}$) and its influence on the external electric fields applied in different orientations. The pink arrows within the schematic diagrams of the films denote the $E_{bi}$, with their sizes indicating the magnitude of $E_{bi}$. The multicolored arrows external to the films represent the composite electric field, formed by the superposition of $E_{bi}$ and the applied electric field ($E_{ap}$), with the length of the arrow proportionally representing the magnitude of the composite field. Spheres of varying colors are indicative of $V_O$ undergoing forward and reverse migration, respectively, with the length of the tails signifying the degree of mobility. (**B**) Illustration of fatigue behaviors for capacitors with various architectures and imprints. (**C**)-(**D**) Positive-up negative-down (PUND) curves and leakage current measurements for HZO devices with diverse structural designs.

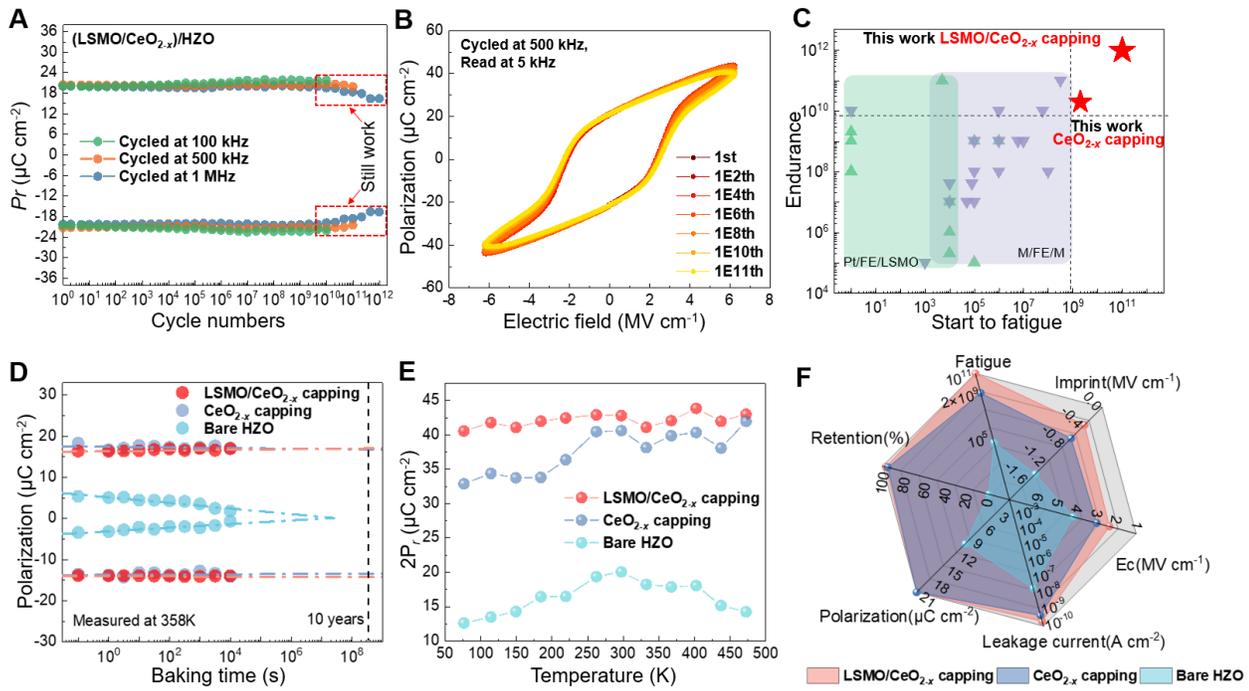

**Fig. 4. Reliability performances for the Pt/LSMO/CeO$_{2-x}$/HZO/LSMO capacitor.** (**A**) Endurance and fatigue behaviors for the LSMO/CeO$_{2-x}$/HZO/LSMO device at different cycling frequencies. (**B**) Polarization-electric field loops of the capacitor at different cycling stage. (**C**) Comparison of the endurance and fatigue manners for different HZO based planar capacitors. (**D**)-(**F**) Retention results, temperature-stability and overall comparisons of electric performances for HZO capacitors with different interfacial designs.

**Supplementary Materials**

Supplementary Text

Supplementary Note1

figs. S1 to S15

References (*59–79*)